\documentclass[sigconf,screen]{acmart}

\usepackage{algorithm} 
\usepackage{algorithmic}  
\usepackage[algo2e]{algorithm2e} 
\usepackage{enumitem}
\usepackage{booktabs}  
\usepackage{threeparttable} 
\usepackage{multirow}
\usepackage{url}
\usepackage{array}
\usepackage{threeparttable}
\usepackage{tikz}
\usepackage{diagbox}
\RequirePackage[normalem]{ulem} 
\RequirePackage{color}\definecolor{RED}{rgb}{1,0,0}\definecolor{BLUE}{rgb}{0,0,1} 
\newcommand*\circled[1]{\tikz[baseline=(char.base)]{
            \node[shape=circle,draw,inner sep=0.7pt] (char) {#1};}}

\AtBeginDocument{%
  \providecommand\BibTeX{{%
    \normalfont B\kern-0.5em{\scshape i\kern-0.25em b}\kern-0.8em\TeX}}}

\copyrightyear{2024} 
\acmYear{2024} 
\setcopyright{acmlicensed}\acmConference[ICSE '24]{2024 IEEE/ACM 46th International Conference on Software Engineering}{April 14--20, 2024}{Lisbon, Portugal}
\acmBooktitle{2024 IEEE/ACM 46th International Conference on Software Engineering (ICSE '24), April 14--20, 2024, Lisbon, Portugal}
\acmPrice{15.00}
\acmDOI{10.1145/3597503.3623313}
\acmISBN{979-8-4007-0217-4/24/04}

\begin{document}

\title{EGFE: End-to-end Grouping of Fragmented Elements in UI Designs with Multimodal Learning}

\author{Liuqing Chen}
\affiliation{%
  \institution{College of Computer Science and Technology, Zhejiang University\\Zhejiang-Singapore Innovation and AI Joint Research Lab}
  \city{Hangzhou}
  \country{China}
}
\authornote{Corresponding author (chenlq@zju.edu.cn).}

\author{Yunnong Chen}
\affiliation{%
  \institution{College of Computer Science and Technology, Zhejiang University}
  \city{Hangzhou}
  \country{China}
}

\author{Shuhong Xiao}
\affiliation{%
  \institution{College of Computer Science and Technology, Zhejiang University}
  \city{Hangzhou}
  \country{China}
}

\author{Yaxuan Song}
\affiliation{%
  \institution{College of Computer Science and Technology, Zhejiang University}
  \city{Hangzhou}
  \country{China}
}

\author{Lingyun Sun}
\affiliation{%
  \institution{College of Computer Science and Technology, Zhejiang University\\Zhejiang-Singapore Innovation and AI Joint Research Lab}
  \city{Hangzhou}
  \country{China}
}

\author{Yankun Zhen}
\affiliation{%
  \institution{Alibaba Group}
  \city{Hangzhou}
  \country{China}
}

\author{Tingting Zhou}
\affiliation{%
  \institution{Alibaba Group}
  \city{Hangzhou}
  \country{China}
}

\author{Yanfang Chang}
\affiliation{%
  \institution{Alibaba Group}
  \city{Hangzhou}
  \country{China}
}

\begin{abstract}
  When translating UI design prototypes to code in industry, automatically generating code from design prototypes can expedite the development of applications and GUI iterations. However, in design prototypes without strict design specifications, UI components may be composed of fragmented elements. Grouping these fragmented elements can greatly improve the readability and maintainability of the generated code. Current methods employ a two-stage strategy that introduces hand-crafted rules to group fragmented elements. Unfortunately, the performance of these methods is not satisfying due to visually overlapped and tiny UI elements. In this study, we propose EGFE, a novel method for automatically End-to-end Grouping Fragmented Elements via UI sequence prediction. To facilitate the UI understanding, we innovatively construct a Transformer encoder to model the relationship between the UI elements with multi-modal representation learning. The evaluation on a dataset of 4606 UI prototypes collected from professional UI designers shows that our method outperforms the state-of-the-art baselines in the precision (by 29.75\%), recall (by 31.07\%), and F1-score (by 30.39\%) at edit distance threshold of 4. In addition, we conduct an empirical study to assess the improvement of the generated front-end code. The results demonstrate the effectiveness of our method on a real software engineering application. Our end-to-end fragmented elements grouping method creates opportunities for improving UI-related software engineering tasks.
\end{abstract}







\maketitle
\section{Introduction}
\label{sec:introduction}

\begin{figure*}[t]
    \centering
    \includegraphics[width=0.98\textwidth]{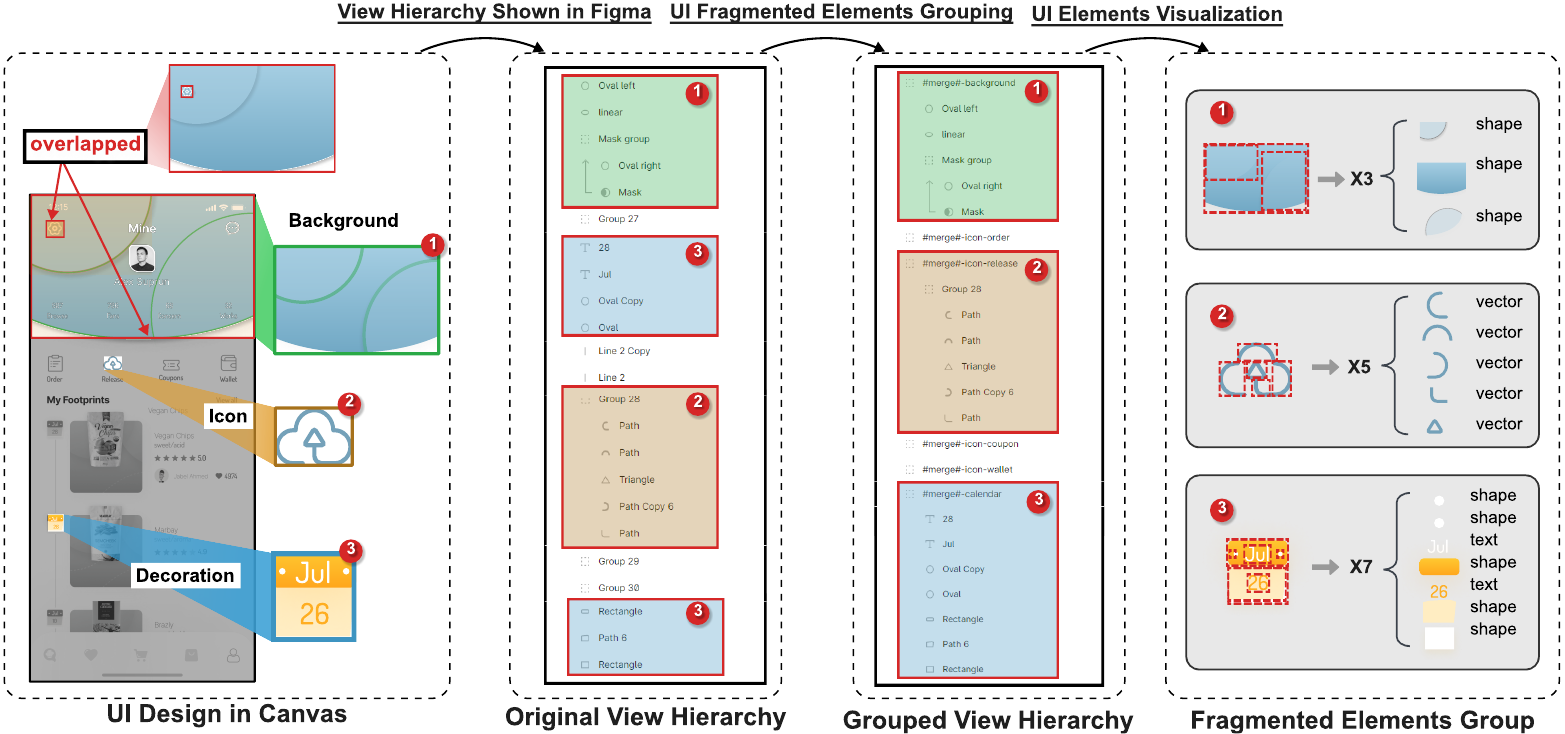}
    \caption{Examples of fragmented elements group (groups are highlighted in solid boxes and fragmented elements are highlighted in dashed boxes)}
    \vspace{-0.11in}
    \label{fig:fragmented_examples}
\end{figure*}

With the rapid progress in electronic devices, there is a growing demand from both companies and individuals for Graphical User Interfaces (GUIs) that provide a superior visual experience. The development of GUI in industry starts from design prototypes\footnote{This work uses "design prototype" and "UI design" interchangeably.} which are produced with design software including Sketch~\cite{sketch} and Figma~\cite{figma}. A design prototype generally consists of multiple UI pages, each featuring a visual effect (UI design in canvas) and a view hierarchy that captures the arrangement of UI elements, as shown in Figure \ref{fig:fragmented_examples}. In industrial settings, front-end developers implement UI design prototypes into code, with a full understanding of the designer's intentions~\cite{autoMachineLearning}. For example, developers can identify discrete objects in the UI that should be instantiated as components on the screen. They categorize these components appropriately based on their intended functionalities and arrange them in a suitable hierarchical structure to ensure correct display on various screen sizes. However, even for skilled developers, this process can be time-consuming and prone to errors. To alleviate the burden on developers, some semi-automated code generation tools, such as Imgcook~\cite{imagecook} and Codefun~\cite{codefun}, utilize design prototypes as inputs to generate code automatically. 
In academia, it is also appealing to propose approaches for generating code from images~\cite{8672718, Beltramelli2018pix2codeGC, autoMachineLearning}, even though the performance is not ready for industrial development. Furthermore, due to the lack of UI view hierarchy, the resulting code often lacks proper structure and accessibility.

In the context of industrial UI prototype design, designers often utilize basic elements to create aesthetic and unique UI components such as backgrounds, icons, and decorations, to enhance the visual appeal. However, due to loose design standards, these elements are typically not combined into a single image view. For example, the ``calendar'' decoration marked as \circled{3} in Figure \ref{fig:fragmented_examples} is formed by 7 basic elements. These individual elements in the view hierarchy that cannot independently convey visual semantics are defined as fragmented elements. During the automated UI-to-code generation process, these fragmented elements can be erroneously interpreted as distinct entities, rather than integral parts of a single, cohesive component. This misinterpretation can cause an escalation in the complexity and redundancy of the generated code. To achieve higher-quality generated code, developers manually group and label UI designs in design software based on visual semantics and the view hierarchy to obtain grouped UI designs, as shown in Figure \ref{fig:fragmented_examples}. However, this grouping process often requires developers to have a thorough understanding of the UI structure and involves iterative trial and error.

In the context of industrial UI-to-code generation, to improve the readability and maintainability of generated code, grouping fragmented elements belonging to the same component in the design prototype is a crucial step, which is referred to as the fragmented element grouping problem~\cite{UILM}. Current approaches~\cite{UILM, uldgnn} for automatically grouping fragmented UI elements employ a two-stage strategy that introduces hand-crafted post-processing to group fragmented elements. However, we found that the two-stage process of identification and grouping can result in cumulative errors, thereby reducing the grouping performance. Furthermore, these approaches are not always robust enough to accurately retrieve all fragmented elements, especially when there is an overlapping issue between background components and foreground components, which is very common in GUIs. For instance, in Figure \ref{fig:fragmented_examples} (UI design in canvas), detecting the bounding boxes for the background component marked as \circled{1} also includes the foreground icon inside, making it difficult to visually separate those overlapped elements in the view hierarchy ~\cite{UILM}. A more intuitive and logically sound approach to address this problem would be to first detect all individual fragmented elements and then merge them into groups based on their spatial position, which is also the pipeline used by most other UI grouping tasks~\cite{ScreenRecognition, Xie2022PsychologicallyInspiredUI}. Although this approach can effectively avoid defects in post-processing, the problem of detecting tiny objects becomes a new bottleneck, as the fragmented elements grouping task deals with the smallest UI element granularity, and many single elements are hard to detect.

In this paper, we propose EGFE, an approach for End-to-end Grouping of Fragmented Elements, which leverages effective UI multimodal feature representation to group UI fragmented elements. EGFE employs a Transformer encoder to model the relationship between UI elements with multi-modal feature representation learning. Specifically, to address the overlapped issue, EGFE followed the idea to process a pixel-based image into a sequence as ViT~\cite{dosovitskiy2020vit}. The element sequence is extracted from the view hierarchy using a depth-first traversal. To address the issue of tiny object detection, we not only consider the vision information of UI elements but also take the attribute information to facilitate UI understanding. To avoid introducing hand-crafted rules, we innovatively transform the detection-based grouping problem into a sequence classification task. EGFE achieves the automatic grouping of fragmented elements as a single stage without introducing any post-processing with hand-crafted rules. To better learn the complex UI context, we use the ResNet model to encode vision features. Moreover, we use the pre-trained BERT~\cite{devlin2019bert} model and fully connected layers to encapsulate the attribute features. To evaluate our approach, we create a dataset containing 4606 UI prototypes designed by professional UI designers. The experimental results demonstrate that our approach outperforms all other baselines. Additionally, we conduct an empirical user study to assess the effectiveness of our method on a real software engineering application. We evaluate the improvement of generated front-end code on four aspects: code availability, modification time, readability, and maintainability. The study shows that EGFE can assist in generating high-quality front-end code at an industrial level.

Our main contributions are outlined as follows:
\begin{itemize}

\item To the best of our knowledge, this is the first attempt to address the grouping of UI fragmented elements with an end-to-end method. The datasets and source code \footnote{\url{https://github.com/test2975/EGFE}} used in our study are publicly accessible to facilitate replication of our research.

\item We propose a novel multi-modal model in EGFE to classify UI fragmented elements and group these elements, and demonstrate its effectiveness in encoding multi-modal features through experiments and an ablation study.

\item A comprehensive evaluation of the performance of EGFE against state-of-the-art baselines, together with an empirical user study, further confirms the effectiveness of EGFE on the industrial aspect.

\end{itemize}
\section{Methodology}
\label{sec:methodology}
\begin{figure*}[tbh]
    \centering
    \includegraphics[width=0.98\textwidth]{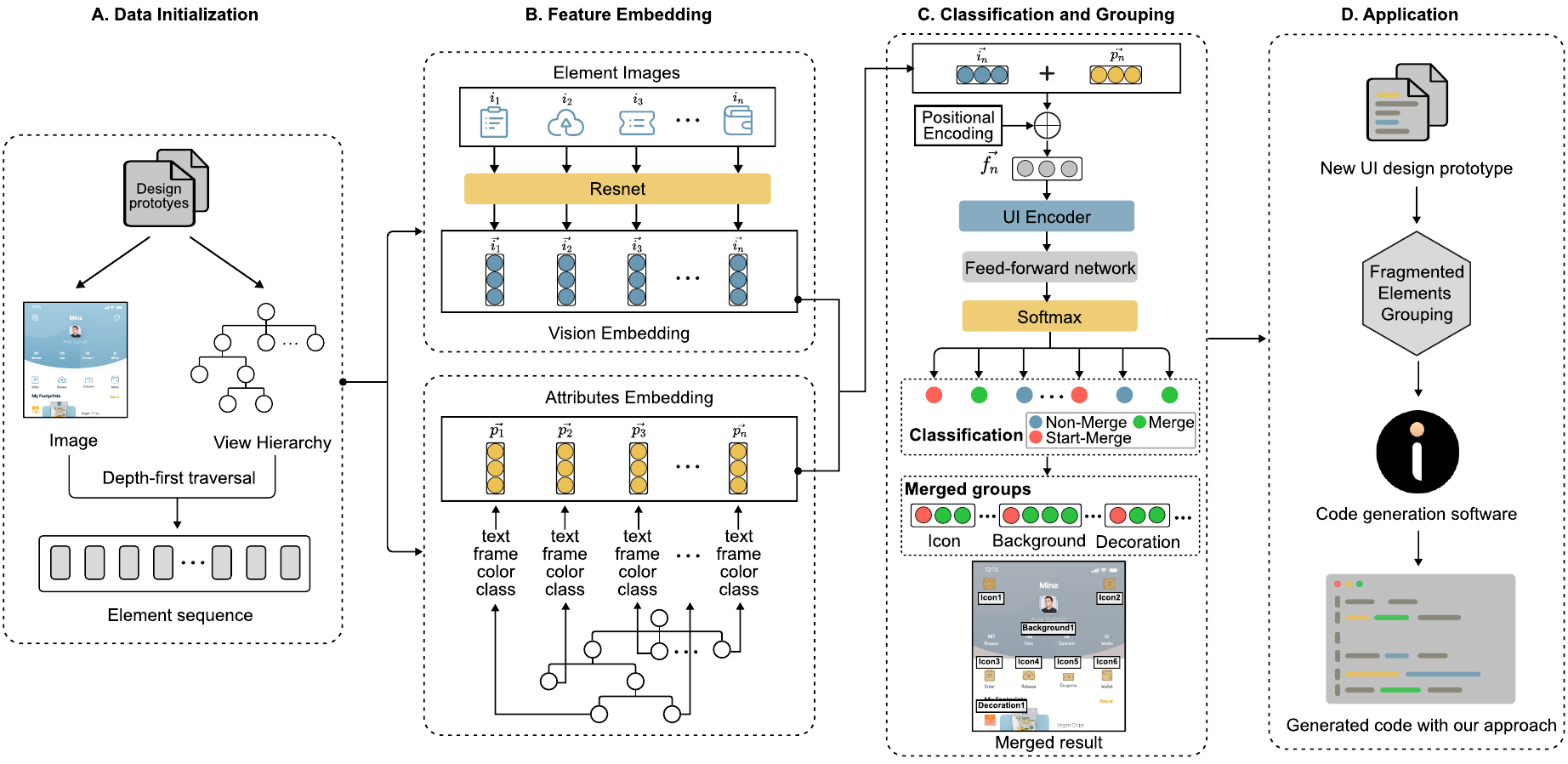}
    \caption{Overview of our approach}
    \vspace{-0.11in}
    \label{fig:overview}
\end{figure*}

Figure \ref{fig:overview} presents the overview of EGFE, which consists of three main steps: (1) \textbf{Data Initialization}, for extracting UI features from the input UI design prototype and generating UI elements sequence; (2) \textbf{Feature Embedding}, for generating deep multi-modal embeddings of UI elements; (3) \textbf{Classification and Grouping}, for classifying UI elements and grouping associated fragmented elements. EGFE is an end-to-end approach. It does not require any ad-hoc or case-specific rules to retrieve UI fragmented elements. Instead, our approach identifies fragmented elements and groups fragmented elements directly. Below, we provide details for each step in EGFE.

\subsection{Data Initialization}
This step is aimed at generating UI element sequences that include both UI element attributes and vision information for a given design prototype, denoted as $E = [e_1, e_2, ..., e_n]$. The predicted elements fall under three categories, namely, \textit{start-merge}, \textit{merge}, and \textit{non-merge}, represented by $e_{sm}$, $e_m$, and $e_{nm}$, respectively. The fragmented element group is a sub-sequence of the prediction sequence with a beginning $start\mbox{-}merge$ node and any number of successive \textit{merge} nodes, denoted as $E_G=[e_{sm},e_m,...,e_m]$.
\subsubsection{UI Design Dataset}
UI design prototypes are high-fidelity GUI prototypes with a view hierarchy that shows how UI elements are constructed and arranged. A view hierarchy is a JSON file with a tree structure where each node in the tree corresponds to an element of a UI design prototype. All elements have meaningful attributes related to the appearance and functionality of UI components, such as content description, frame (position and size), color, and class. When preparing the experimental dataset, as shown in Figure \ref{fig:fragmented_examples}, fragmented components are classified into three categories based on their spatial size and visual effect including icon, decoration, and background. These components are often highly customized, reflecting the aesthetics and characteristics of the user interface. To train and evaluate our approach, we collect 4606 design prototypes created on Sketch~\cite{sketch}. These UI design prototypes are designed by professional designers for various kinds of apps such as shopping, travel, and communication. To obtain vision information of UI elements, we use Sketch command-line interface~\cite{sketchcli} to export element images. We assign each element a label: \textit{start-merge}, \textit{merge}, and \textit{non-merge}. Specifically, we assign the label \textit{start-merge} to the first traversed element of each merged group. The remaining elements within a merged group are labeled as \textit{merge}, while the elements outside of merged groups are labeled as \textit{non-merge}.

\subsubsection{Annotation Process} 

As data annotation has become more difficult due to the need for large-scale data and the lack of adequately labeled data, crowdsourcing~\cite{su2012crowdsourcing} has gained considerable attention within the computer vision community. In our study, we employed crowdsourcing to comprehensively annotate the EGFE dataset, recruiting 18 professional front-end developers with over five years of experience each. The annotators received detailed instructions and examples of fragmented element groups, and each annotator was required to annotate approximately 250 design prototypes. Following the initial annotation, two authors of this paper, a researcher and a senior front-end engineer, meticulously assessed the quality of all design prototypes and identified any annotation errors. Annotators were then instructed to revise their annotations based on our feedback and re-annotate any UI designs that did not meet our quality standards. During the labeling process, annotators need to identify the associated fragmented elements and group them, assigning the fragmented elements group a ``\#merge\#'' label. The view hierarchy of fragmented element groups is illustrated in Figure \ref{fig:fragmented_examples} (Grouped view Hierarchy). On average, a duration of approximately 10 minutes was dedicated to each design prototype to ensure that the labeling quality met our standards.

\subsection{UI Elements Embedding}

\subsubsection{Image Embedding}

Most of the current research~\cite{bai2021uibert, 3411764.3445762, 10.1145/3491102.3502042} treats UI images as image patches based on the bounding boxes denoted in a view hierarchy. The image patches contain their surrounding information. Inspired by the Gestalt principle~\cite{Gestalt, Xie2022PsychologicallyInspiredUI}, they suggest that elements are perceptually grouped if they are similar to each other. The different surrounding information may destroy the semantics of the element itself. For example, in the background component shown in Figure \ref{fig:fragmented_examples}, splitting image patches of background elements would result in including the overlapping UI elements. This redundant vision information would hinder the identification of individual background elements. Therefore, we directly use element images instead of encoding the entire image patches. To reduce the image processing time, we normalize the element images to a fixed size of $3\times64\times64$ before training. We use Resnet-50~\cite{7780459} of which the last fully connected layer is replaced by $d$ dimension to encode the pixel information of elements, where $d$ is the dimension size that is 256 in our case. The output is a two-dimensional tensor $i\in \mathbb{R}^{L\times d}$, with each row serving as the visual encoding of an element.

\subsubsection{Attributes Embedding} 
The UI elements in Design prototypes contain multiple attributes, including text descriptions, class, color, and frame. To encode each element as a dense vector, we embed its attributes separately and then combine these embeddings. The element text description is a brief description of the functionality of this UI element provided by the designer. If the delivery is badly organized, the developer can use the text description as a prompt to group related fragmented elements together. For instance, shape elements typically contain keywords like ``path'' and ``rectangle'' while elements representing symbols are often named based on their functionality, such as ``status bars''. To achieve an adaptable representation of the element text, we normalize all the text with lower letters, and a specific tokenizer with pre-trained BERT~\cite{devlin2019bert} is adopted. We embed each text into a tensor with a length of 32, with each value denoting the index of the word in the pre-trained corpus. Two-sided padding (with 0) and truncation strategies are used to eliminate the effect of text length differences. Every text is represented as a $32\times d$ tensor. To ensure synchronization with the image features, we compress the first dimension by summing up each column. This results in a text embedding $p_{text} \in \mathbb{R}^{L\times d}$.

The element color item $[R, G, B, A]$ in our dataset denotes each element's background color. Note that some elements do not include such background color information, and we just leave them as zero vectors. We normalize the color between 0 and 1. The bounding box $[x, y, w, h]$ defines the frame of UI elements. Considering the different sizes of the design prototypes, we convert the data pair into $[x, y, x+w, y+h]$ format and then apply min-max normalization to scale all the bounding boxes into the same range. We adopt a similar way to process the color embedding $p_{color}$ and frame embedding $p_{frame}$ by using fully connected layers, and both of their output shapes are $L\times d$.

The class indicates the predefined type in the UI design prototype, such as the oval, rectangle, or text. It is a useful attribute that helps EGFE learn specific patterns. For example, the text element is usually more unlikely to be part of the fragmented elements, as it makes no sense to separate some text with coherent semantics into several parts in the design stage. Such prior information from the class is valuable for our prediction. Instead of processing the class of each element as a one-hot vector, we use an embedding layer, resulting in an output feature shape of $p_{class} $ as $L\times d$. By obtaining an embedding of the class, we achieve a more pronounced impact at the fusion stage. In contrast, a one-hot vector is too sparse to fulfill this expectation, as most of its values are 0.

\subsection{Classification and Grouping}
We construct a UI encoder to capture the relationship between elements by following the structure of the DETR Encoder~\cite{carion2020end}. Overall, our model uses the UI encoder to encode the entire embedding and a two-layer Feed-Forward Network (FFN) with a ReLU activation function to predict the type of each element. Specifically, given a element sequence $E = [e_1, e_2, ..., e_n]$, where $n$ is the number of elements in the sequence. The image embedder first embeds each image of the elements into a $d$ dimensional vision embedding $\vec{i}$. Then, the attributes embedder embeds attributes into $d$ dimensional embeddings (i.e., $\vec{p}_{frame},\vec{p}_{class},\vec{p}_{text},\vec{p}_{color}$). The final embeddings are as follows:
\begin{equation}
\overrightarrow{e_i} = \vec{i}+\vec{p}_{frame}+\vec{p}_{class}+\vec{p}_{text}+\vec{p}_{color}
\end{equation}
We use the $sine$ and $cosine$ functions of different frequencies to compute the position encoding $PE_i$:
\begin{align}
    PE_{i,2j} &= \sin{\big(j/{10000}^{2j/d}\big)} \\
    PE_{i,2j+1} &= \cos{\big(j/{10000}^{2j/d}\big)}
\end{align}
where $i$ is the position of the element and $j$ denotes the $j$-th dimension of the embedding vector.
Then, the final embeddings with positional encoding are as follows:
\begin{equation}
\overrightarrow{f_i} = \overrightarrow{e_i}+PE_i
\end{equation}
Then, the UI encoder inputs the sequence of element embeddings into $N$ identical Transformer encoder blocks to calculate the hidden states of the element. For the $i^{th}$ block of the UI encoder, suppose that the input is $H_{i-1}$, and the output $H_{i}$ is calculated as follows:
\begin{align}
H_{i,1} &= {\rm LayerNorm}\bigg(H_{i-1} + {\rm MHAtt}(H_{i-1})\bigg) \label{eq:first_layer} \\
H_{i} &= {\rm LayerNorm}\bigg(H_{i,1} + {\rm FFN} (H_{i,1})\bigg) \label{eq:last_layer}
\end{align}
where $H_{i,1}$ is the hidden states of the first sub-layer and $MHAtt$ is the multi-head attention layers. Initially, the element embedding vectors $[\overrightarrow{f_1}, \overrightarrow{f_2}, ..., \overrightarrow{f_n]}$ are fed into the first block, and the $N^{th}$ block outputs the final hidden states of the input element $H=[h_1, h_2,...,h_n]$. Then, it uses a two-layer FFN to map the features into the classification representation $c^k$.
\begin{align}
c^{k} &= {\rm FFN}(c^{k}) = {\rm ReLU}(c^{k}\cdot W_1 + b_1)\cdot W_2 + b_2 
\end{align}
With a Softmax function, the outputs of the ${\rm Softmax}(c^{k})$ is a distribution of our element prediction probability $P_z\in \mathbb{R}^{n\times 3}$ with the $i^{th}$ row represents the classification probability of the $i^{th}$ element. Finally, we use the cross-entropy metric to calculate the loss between each prediction and ground truth.

The element sequence we obtained has eliminated the hierarchical bias that may be caused by various design patterns. As shown in Figure \ref{fig:overview}, our approach assigns a label to each element. Any potential groups of fragmented elements are naturally grouped as sub-sequences without rule-based grouping algorithms. We start grouping the element with a \textit{start\mbox{-}merge} label and continue to group elements with a \textit{merge} label until traversed another \textit{start\mbox{-}merge} element. During traversal, the elements with \textit{non\mbox{-}merge} label are filtered. 

\subsection{Application}
Competitive GUI development typically requires a large team of designers and engineers with specialized domain expertise. The UI design and code implementation process often involves multiple iterations to ensure that end-users experience the UI as intended, resulting in a time-consuming and error-prone process~\cite{moran2018automated}. Prior methods based on image-captioning~\cite{Beltramelli2018pix2codeGC, autoMachineLearning} have attempted to automate the code development process to address this repetition. However, the generated code is highly redundant for repetitive GUI blocks as it lacks awareness of the GUI structure during code generation. More recent industrial solutions, such as Imgcook~\cite{imagecook}, have addressed this limitation by utilizing design prototypes containing UI metadata to avoid code redundancy and structural loss. However, this approach lacks awareness of fragmented elements, resulting in the generation of redundant image containers. This requires developers to invest additional effort in revising the code. Our approach can group fragmented elements and generate less redundant and more reusable GUI code. In Section \ref{sec:human_evaluation}, we conduct an empirical user study to demonstrate the effectiveness of our approach in this downstream automation software engineering task.
\section{Experiments}
\label{sec:exp}
According to our research purpose, the evaluation aims to answer the following research questions:

\textbf{RQ1:(Elements Classification Performance)} How does the EGFE perform compared to the state-of-the-art image classification baselines?

\textbf{RQ2:(Fragmented Elements Grouping Performance)} What's the performance of EGFE on fragmented elements grouping?

\textbf{RQ3:(Multimodal UI Features Performance)} How does each individual feature in EGFE improve the overall performance?

To fully understand the performance of our sequence-based approach, we compare it with the state-of-the-art image classification method in RQ1. We examine the grouping performance of EGFE and compare it with state-of-the-art baselines in RQ2. As we observe that fusing multimodal features improves the classification performance, we figure out the influence of each feature in RQ3.

\subsection{Dataset}

As described in \ref{sec:methodology}.1, we generate element sequences from manually labeled design prototypes. During the data labeling process, the original order of elements may be disrupted. Therefore, we generate element sequences based on the original design prototypes and the element UUIDs obtained from the labeled design prototypes. Regarding our training-testing data split, the split was performed file-wise, i.e., prototypes from the same design file are not shared among the three splits. This is to avoid information leakage because prototypes from the same package might have similar appearances. We split the design prototypes in the dataset, completing the $8:1:1$ split for the training, validation, and test sets, respectively. The resulting split has 3686 design prototypes in the training dataset, 460 design prototypes in the validation dataset, and 460 design prototypes in the testing dataset. Table \ref{tab:source} shows the number of elements in three sets. The total ratio of merged versus non-merged elements is approximately $1:8$. 

\begin{table}[thp]
  \vspace{-0.11in}
  \caption{Dataset statistics}
  \vspace{-0.04in}
  \centering
  \label{tab:source}
  \begin{tabular}{lccccc}
    \toprule
    \textbf{Split} &\textbf{S} &\textbf{M} &\textbf{N} &\textbf{Tiny} &\textbf{Overlap}\\
    \midrule
    Training      & 15,247  &23,851 & 287,513 & 123,734 & 1,592 \\
    Validation    & 1,258 &2,263  & 35,939 & 25,645 & 231 \\
    Test          &   2,554    &3,700  &30,480 &28,982 & 412 \\
    \midrule
    Total &19,059 &29,814 &353,932 &178,361 & 2,235\\
    \bottomrule
  \end{tabular}
\begin{tablenotes}
    \footnotesize
    \item S, M, N denotes \textit{Start-merge}, \textit{Merge} and \textit{Non-merge}, respectively.
\end{tablenotes}
\vspace{-0.25in}
\end{table}

\subsection{Implementation Details}
We use a ResNet-50 model pretrained on ImageNet as an element image encoder. The text embeddings are initialized with a pretrained BERT encoder. The hidden dimension $d$ is set to 256 for all the embeddings. For the Transformer Encoder, we use $N=6$ Transformer layers and 8 attention heads with MLP dimension 2048 and query/key/value dimension 256. We train our models with a mini-batch of 8 samples for 300 epochs using the Adam optimizer with a $learning~rate=1e-4$. To avoid over-fitting, we set $dropout=0.2$, and adopt the $L2\mbox{-}regularization$ with $\lambda=1e-5$. The learning rate is divided by 10 times after 200 epochs. To describe the multi-class classification task, we adopt the cross-entropy loss to train our EGFE. We employ class weights to alleviate the issue of long-tail distribution in the data. Our approach is implemented based on the Pytorch~\cite{pytorch} framework. The experimental environment is a desktop computer equipped with an NVIDIA GeForce RTX 3090 GPU, intel core i7 CPU, and 32GB RAM, running on Ubuntu OS and it cost 12 hours to converge.

\subsection{Baselines}
To the best of our knowledge, only a few works focus on specific element grouping tasks, and no existing work in this area follows an end-to-end pipeline to ours. In this case, we compare the classification and group performance of EGFE with the state-of-the-art baselines, respectively. 

\textbf{For classification performance comparison}, we choose the following baselines: 

\textbf{EfficientNet}~\cite{tan2019efficientnet} is a CNN-based method that uses a simple yet highly effective compound coefficient to scale up CNNs in a more structured manner. Recently, it has been a frequently employed baseline in image classification tasks.

\textbf{Vision Transformer}~\cite{dosovitskiy2020vit} Known as ViT succeeded in using a full transformer to outperform previous works based on convolutional networks in the vision field. 

\textbf{Swin Transformer}~\cite{liu2021swin} is a hierarchical Transformer whose representation is computed with shifted windows. It achieves strong performance on many vision downstream tasks. 

We trained these image classification models using the same element image data as EGFE from the initial state. The trained models are capable of performing the same classification tasks as EGFE.

\textbf{For grouping performance comparison}, we choose the following baselines:

\textbf{UILM}~\cite{UILM} is an object detection-based method for grouping fragmented elements in UI designs. Initially, it uses a novel object detector to detect UI component regions that consist of fragmented elements. Then, it proposes a rule-based algorithm to retrieve the fragmented elements within the bounding boxes.

\textbf{ULDGNN}~\cite{uldgnn}, a method grouping fragmented elements based on UI view hierarchy, transforms the layout of each UI design prototype into a graph with each node representing a UI element. ULDGNN employs a self-attention-based graph neural network to identify whether an element is fragmented. Subsequently, a rule-based algorithm is implemented to group fragmented elements.

\textbf{UIED}~\cite{10.1145/3368089.3409691} is a hybrid object detection approach for UI element detection. It is demonstrated to outperform YOLO v3, FasterRCNN, and Centernet. UIED first employs traditional computer vision methods to detect the boundaries of elements (referred to as components in this paper) and then uses CNN to classify these components. 

We replicated the models of UILM and ULDGNN and trained them on our dataset from the initial state. For UIED, we follow the detection approach of UIED and retrain the region classifier of UIED with our dataset to enable it to recognize component regions containing fragmented elements. Similar to UILM, the rule-based grouping algorithm is utilized to identify fragmented elements within the regions.

\subsection{Evaluation Metrics}

\textbf{UI elements classification.} To evaluate the classification performance of EGFE and image classification baselines, we use precision, recall, and F1-score to measure the performance of element classification. 

\textbf{UI elements grouping.} To evaluate the grouping performance of EGFE and state-of-the-art baselines, we generate the ground-truth UI element sequence as a string through depth-first traversal, based on the design prototype's view hierarchy. As described in Section \ref{sec:methodology}, only UI elements with \textit{non-merge} labels are filtered, resulting in ground-truth groups consisting solely of \textit{merge} and \textit{tart-merge} types. To enable comparison, the predicted merged groups generated by our approach and baselines are outputted in the same format. The Levenshtein edit distance is computed between the ground-truth sequence and the predicted sequence, revealing the specific mismatches between the two groups and aiding in the analysis of grouping errors. To determine the optimal matching between the ground-truth group string and the predicted merged group string, we minimize the overall edit distance among all candidate matches. If a merged group corresponds to a ground-truth group, it is classified as a true positive (TP); otherwise, it is a false positive (FP). Any ground-truth groups that do not correspond to any merged group are classified as false negatives (FN). Based on the matching results, we compute (1) precision $(TP/(TP+FP))$ (2) recall $(TP/(TP+FN))$, and (3) F1-score $((2*precision*recall) / (precision+recall))$

\subsection{RQ1: Performance of UI Elements Classification}

The results in Table \ref{tab:classification} show that EGFE outperforms all baselines in every situation. Our method outperforms the state-of-the-art image classification baseline (Swin Transformer) in terms of precision, recall, and F1-score, with improvements of 25.5\%, 22.7\%, and 24.0\%, respectively. EGFE$^{image}$ means using image features only. The performance of EGFE$^{image}$ is also better than that of other baselines. This is because image classification models focus on classifying individual elements without considering the relationships between them. In contrast, EGFE$^{image}$ employs a Transformer encoder to model the relationships between image features of elements, leading to an improvement in performance. Besides, regardless of whether the image classification model is based on CNN or Transformer, relying solely on pixel information makes it challenging to fully understand the UI design prototype, resulting in lower performance. By employing multi-modal representation learning, we enable EGFE to better understand design prototypes, consequently yielding significant performance advancements. This suggests that improving the representation of multi-modal features is a crucial aspect of improving model performance. The result in Table \ref{tab:acc} also shows that our approach has good performance in three types, especially in the \textit{non-merge} type. Since there are fewer fragmented elements in the design prototypes than ordinary elements, more training examples of fragmented object types would potentially improve the model performance.

\begin{table}[htp]
\centering
\caption{Comparison of element classification accuracy}
\vspace{-0.04in}
\label{tab:classification}
  \begin{tabular}{lccc}
    \toprule
    \textbf{Method} &\textbf{Precision} &\textbf{Recall} &\textbf{F1}\\
    \midrule
    EfficientNet & 0.568 & 0.580 & 0.574 \\
    Vision Transformer & 0.625 & 0.638 & 0.631 \\
    Swin Transformer & 0.675 & 0.682 & 0.679 \\
    EGFE$^{image}$ & 0.769 & 0.786 & 0.777 \\
    EGFE &\textbf{0.847} & \textbf{0.837} & \textbf{0.842} \\

    \bottomrule
  \end{tabular}
  \vspace{-0.11in}
\end{table}

\begin{table}[htp]
\centering
\caption{The accuracy of each type for EGFE}
\vspace{-0.04in}
\label{tab:acc}
  \begin{tabular}{lccc}
    \toprule 
    \textbf{Element Type} &\textbf{Precision} &\textbf{Recall} &\textbf{F1}\\
    \midrule
    Non-merge & 0.980 & 0.956  & 0.968 \\
    Start-merge & 0.764  & 0.772  & 0.767  \\
    Merge & 0.797 & 0.783 & 0.790  \\
    \bottomrule
  \end{tabular}
\vspace{-0.11in}
\end{table}

\subsection{RQ2: Performance of UI Elements Grouping}
\begin{figure*}[htp]
  \centering
    \includegraphics[width=0.95\textwidth]{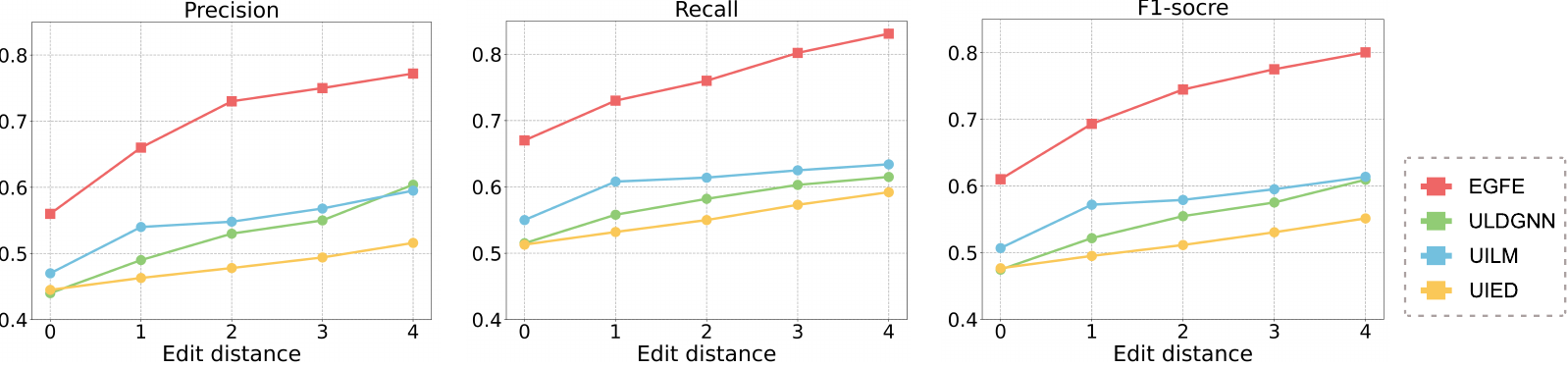}
  \caption{Grouping performance at different edit distance thresholds (0-4)}
  \label{fig:exp_grouping}
\end{figure*}

The results in Figure \ref{fig:exp_grouping} show that EGFE outperforms baselines in five edit distance thresholds (0-4). The distance 0 means the two groups of a ground-truth merged group and a predicted merged group have the perfect match. The distance 4 means the unmatched elements in the two groups are no more than 4. As shown in Figure \ref{fig:exp_grouping}, as the distance threshold increases, which means the matching criterion relaxes, the precision, recall, and F1-score keep increasing to 0.772, 0.831, and 0.80 at threshold 4. Our method outperforms the state-of-the-art baselines in the precision (by 29.75\%), recall (by 31.07\%), and F1-score (by 30.39\%) at the edit distance threshold of 4. It indicates that our approach can classify and group UI elements more precisely than all baselines. Due to the limitations of traditional computer vision techniques in generalizing to complex UI designs, the performance of UIED in detecting component boundaries is not satisfied, which subsequently affects the classification of these components. Consequently, the detection and identification of components that contain fragmented elements are also impacted, resulting in unsatisfactory performance. Despite the progress made by UILM in enhancing detection performance, it still faces challenges in accurately identifying the elements within the detected bounding boxes. Our EGFE also achieves a noticeable improvement over the two-stage graph-based approach ULDGNN. Our approach leverages the accessible layer element information from design prototypes, thereby eliminating the need for element localization. By employing a multimodal Transformer and multi-class classifier, it achieves end-to-end recognition and grouping of fragmented elements.

Table \ref{tab:performance_tiny} demonstrates the superior performance of EGFE compared to baselines in grouping tiny and overlapping elements. For tiny elements, we follow the COCO format~\cite{lin2014microsoft}, which considers an area smaller than $32\times32$ pixels. For elements with large aspect ratios in the designs, we require both the length and width of tiny elements must be less than $32$ pixels. Regarding overlapping elements, we automatically select samples from the test set that have overlapping bounding boxes and calculate precision, recall, and F1-score for these samples. The results show that EGFE outperforms the best baseline (UILM) by 41.7\%, 30.6\%, and 36.2\% in terms of precision, recall, and F1-score for overlapping elements. Our model also outperforms the best baseline by 19.9\%, 16.0\%, and 18.0\% in terms of precision, recall, and F1-score for tiny elements. This is primarily due to the fact that EGFE not only identifies elements from a visual perspective but also incorporates UI attribute information from the design metadata.

\begin{table}[htp]
\centering
\vspace{-0.04in}
\caption{Performance of grouping tiny and overlapping elements \textbf{(edit distance $\le$ 1)}}
\label{tab:performance_tiny}
 \begin{tabular}{lcccccc}
  \toprule
  & \multicolumn{3}{c}{\textbf{Tiny Elements}} & \multicolumn{3}{c}{\textbf{Overlapping Elements}} \\
  \midrule
  \textbf{Method} & \textbf{prec.} & \textbf{rec.} & \textbf{F1} & \textbf{prec.} & \textbf{rec.} & \textbf{F1}\\
  \midrule

ULDGNN & 0.466 & 0.537 & 0.499 & 0.453 & 0.478 & 0.465\\
UIED & 0.420 & 0.497 & 0.455 & 0.209 & 0.235 & 0.221 \\
UILM & 0.523 & 0.582 & 0.551 & 0.460 & 0.487 & 0.473 \\
EGFE & \textbf{0.627} & \textbf{0.675} & \textbf{0.650} & \textbf{0.652} & \textbf{0.636} & \textbf{0.644} \\

  \bottomrule
 \end{tabular}
\vspace{-0.04in}
\end{table}

\begin{figure}[htp]
  \centering
  \includegraphics[width=0.46\textwidth]{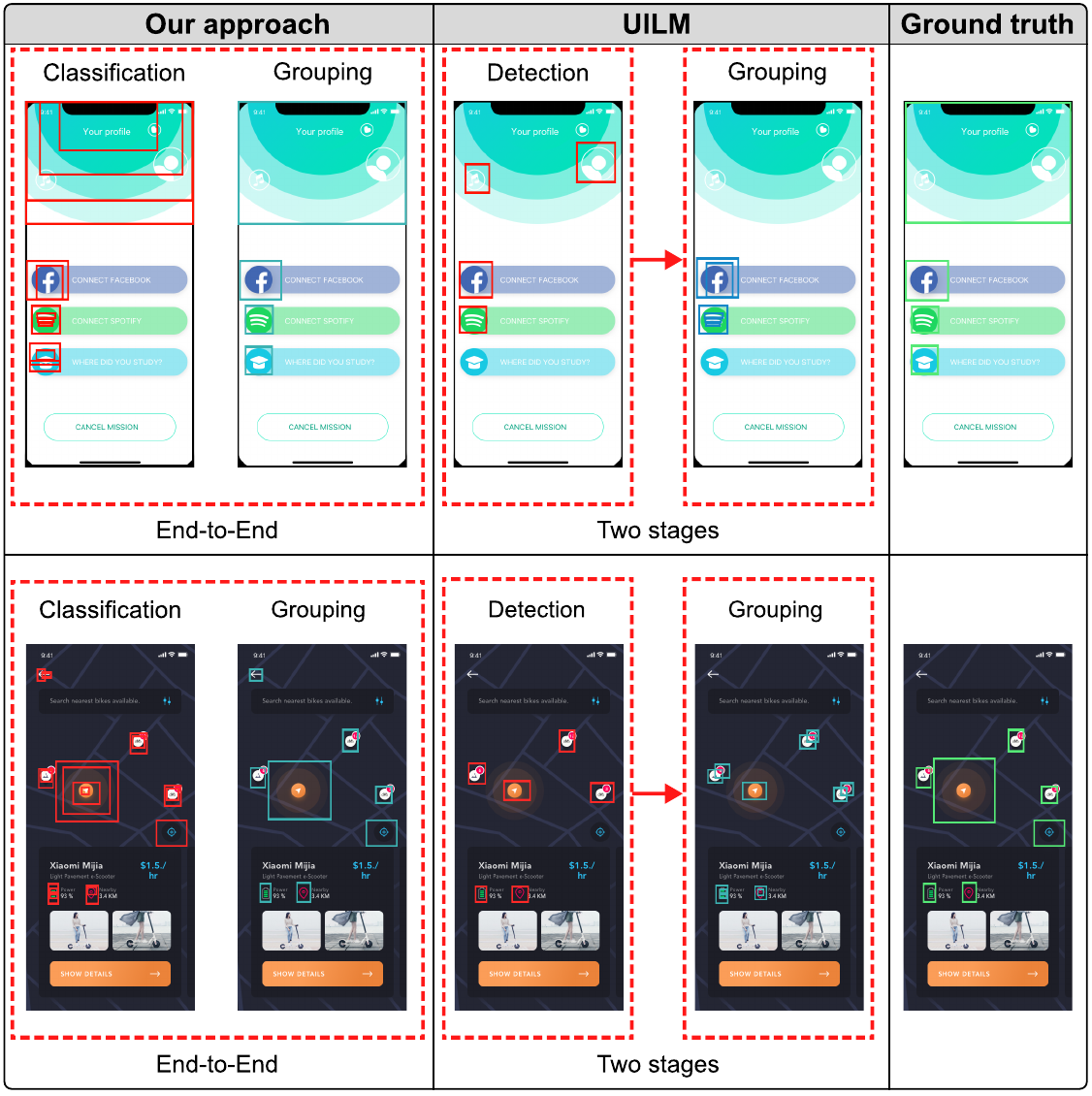}
  \caption{Examples of end-to-end approach EGFE and two stages approach UILM (in EGFE, red box - predicted elements, blue box - merged group and in UILM, red box - predicted bounding boxes, blue box - retrieved elements, green box - ground-truth groups)}
  \label{fig:exp_pipeline}
  \vspace{-0.11in}
\end{figure}

We also present examples in Figure \ref{fig:exp_pipeline} for intuitive comparison. Our approach can not only accurately recognize the complex merged group of background elements, but it can also reduce false positive predictions with powerful multimodal learning ability. For example, in the first row of Figure \ref{fig:exp_pipeline}, the background object composed of multi-semicircular elements is not detected by the object detection method, and it also incorrectly recognizes two complete UI icons on the background object. However, our approach still robustly recognizes the groups of fragmented elements.

\begin{figure}[thp]
  \centering
  \includegraphics[width=0.45\textwidth]{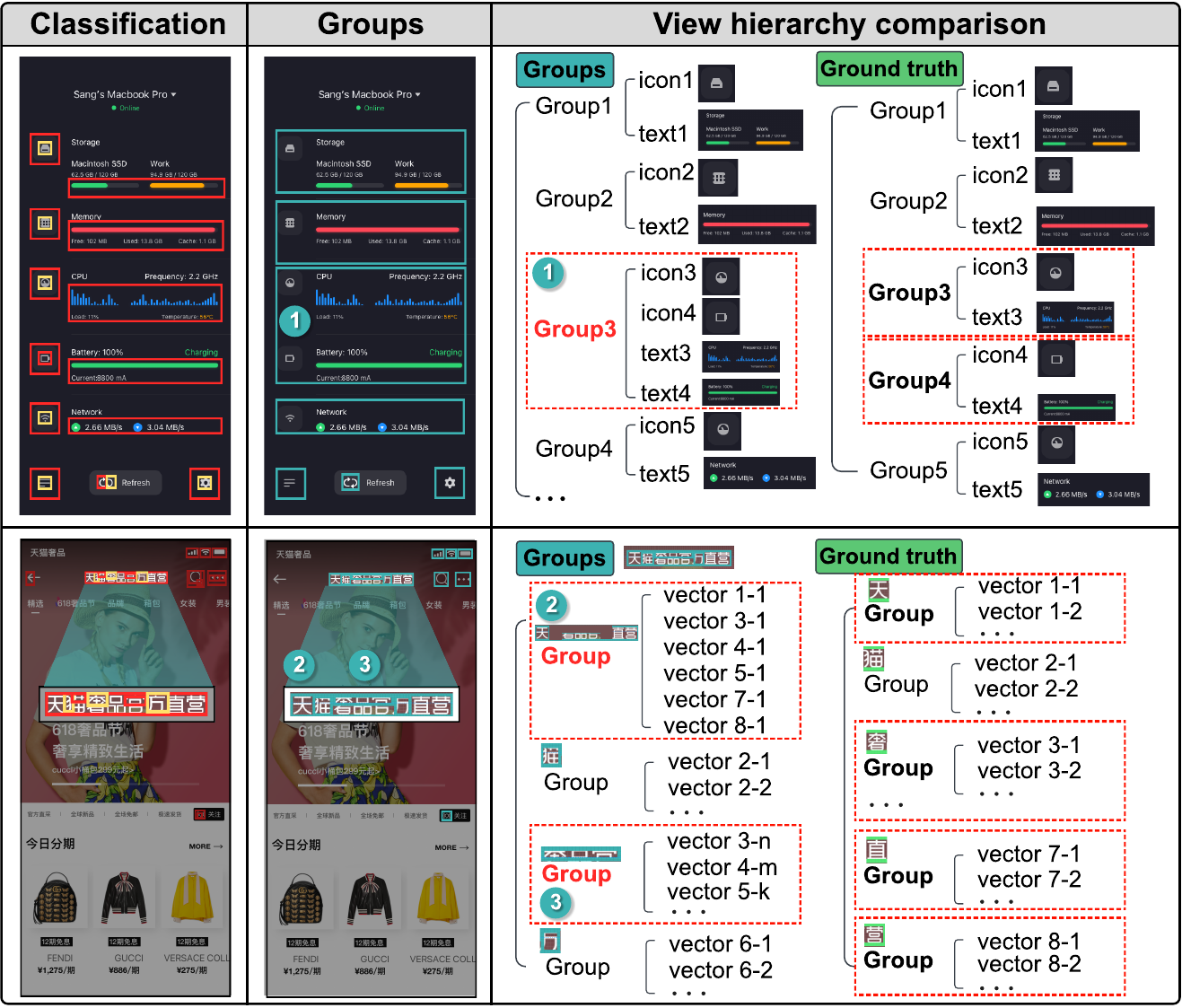}
  \caption{Typical causes of grouping mistakes (in UI screenshots, yellow box - \textit{start-merge} elements, red box - \textit{merge} elements, blue box - merged groups and in the view hierarchies, red dashed box - incorrect groupings)}
  \label{fig:failures}
  \vspace{-0.11in}
\end{figure}

We present two typical failure cases to highlight the potential improvements of our approach. As demonstrated in Figure \ref{fig:failures}, the primary challenge of the end-to-end grouping approach is detecting and classifying elements accurately. In the first row, our approach successfully identifies all fragmented elements on the GUI, but due to misclassification, it groups nearby widgets as a single component marked as \circled{1}. This issue could be addressed by reducing data imbalance or implementing a weighted training strategy. In the second row, our approach segments the artistic font component into several groups due to missed element detection. This is because artistic fonts are composed of many ``shape'' elements, which results in misunderstandings by the model. This issue could be alleviated by introducing an OCR tool to predict the artistic font component. This approach can also be considered end-to-end as it does not involve any additional rules.

\subsection{RQ3: Ablation Experiment on Multimodal Features}

To investigate the effectiveness of each UI feature, we discard each at a time. The experimental results of the ablation study are shown in Table \ref{tab:ablation}. As arranged in the first column, EGFE represents our ensemble performance using image and element properties (i.e., text, class, frame, and color), and each row below represents the result of a missing feature. For example, EGFE w/o text denotes the model without the attribute of text description. Table \ref{tab:ablation} shows that each feature contributes to the overall performance. The results that the removal of the image feature leads to a dramatic decrease in both macro (by 18.93\%) and weighted (by 9.09\%) F1 scores, which has strongly indicated the element image feature is essential information that contributes to the high performance of our EGFE. We can also observe that removing the text description feature will lead to a performance decline in the macro F1 (by 7.54\%), and weighted F1 (by 4.95\%).

\begin{table}[htp]
\centering
\vspace{-0.11in}
\caption{Ablation results}
\label{tab:ablation}
\resizebox{0.48\textwidth}{!}{
 \begin{tabular}{lcccccc}
  \toprule
  & \multicolumn{3}{c}{\textbf{Macro Average}} & \multicolumn{3}{c}{\textbf{Weighted Average}} \\
  
  \midrule
  
  \textbf{Method} & \textbf{prec.} & \textbf{rec.} & \textbf{F1} & \textbf{prec.} & \textbf{rec.} & \textbf{F1}\\
  \midrule

\textbf{EGFE} & \textbf{0.847} & \textbf{0.837} & \textbf{0.842} & \textbf{0.919} & \textbf{0.904} & \textbf{0.912} \\

 w/o text & 0.794 & 0.772 & 0.783 & 0.879 & 0.859 & 0.869 \\
 w/o color & 0.828 & 0.811 & 0.820 & 0.910 & 0.897 & 0.903 \\
 w/o class & 0.819 & 0.793 & 0.806 & 0.896 & 0.885 & 0.891 \\
 w/o frame & 0.827 & 0.811 & 0.819 & 0.906 & 0.895 & 0.901 \\
 w/o image & 0.716 & 0.700 & 0.708 & 0.836 & 0.836 & 0.836 \\

  \bottomrule
 \end{tabular}
 }
\vspace{-0.11in}
\end{table}

\section{Empirical study on code generation}
\label{sec:human_evaluation}
Our fragmented grouping method fills an important gap in UI understanding. The experiments above demonstrate the performance of our approach with other baselines. However, the satisfactoriness of the predicted merged group can be subjective to different users or developers. Therefore, we choose GUI-to-code generation, an innovative application in the software engineering domain, to demonstrate the effectiveness of our method. We perform an empirical human evaluation to further assess the quality of code improvement by applying EGFE. We also investigate the feedback from developers.

\subsection{Procedure}

We conduct a comprehensive investigation to assess the quality of the generated code with our approach. To carry out this experiment, we utilize vscode~\cite{vscode} as the development environment for participants and Imgcook~\cite{imagecook} to generate the React-format code with and without fragmented element grouping. We recruit ten participants, all of whom possess over three years of programming experience in front-end development using the JavaScript framework, and can be considered senior front-end code developers capable of evaluating the quality of our generated code. We randomly selected ten UI design prototypes for each of the three typical categories of UI design prototypes, namely travel, communication, and shopping. For comparison purposes, we use the original generated code from Imgcook as the baseline, which does not group fragmented elements. The final number of code snippets generated by Imgcook is 60 in total, comprising 30 code snippets from the merged UI prototypes by EGFE and 30 code snippets from the non-merged ones.

Initially, we provide generated code snippets to each participant with and without applying EGFE. The participants then revise each code snippet with rendered GUI until they believe the code structure meets the production standards. It is important to note that they are not aware of which code snippets are from which method, and all of them evaluate the code individually without any discussion. After the experiment, each participant rates each generated code based on four aspects stated in advance: (1) \textbf{Readability} reflects the difficulty of understanding the generated code from the perspective of structure; (2) \textbf{Maintainability} reflects the difficulty of maintaining the generated code for the next requirement; 3) \textbf{Code Availability} reflects how much code is available for prototyping; 4) \textbf{Code Modification Time} reflects how much time is required to adjust the code to actual production standards. For the first two aspects, participants should rate scores, ranging from 1 to 5 (1 for poor and 5 for excellent). We calculate the code availability as follows:
\begin{equation}
    code\ availability = 1-\frac{lines\ of\ code\ changes}{total\ lines\ of\ code}
\end{equation}

We divide the score range [0.80, 0.85, 0.90, 0.95] into intervals of 5\% code changes and assign a score to each interval from 1 to 5. We use this mapping to calculate the code availability score. For the code modification time, we record the time taken by each participant to adjust the generated code to actual production standards. We set the interval for this aspect as 5 minutes, where a score of 5 is assigned to those who complete the task within five minutes, and a score of 1 is assigned to those who take more than 20 minutes.

\subsection{Results}

Table \ref{tab:userstudy} shows the statistic results. Overall, EGFE is better than the baseline in three types of design prototypes. The average score for code availability, modification time, readability, and maintainability are 3.41, 3.46, 4.00, and 3.61, respectively. In terms of code availability, modification time, readability, and maintainability, which are four metrics for evaluating code quality, EGFE has increased by 46.35\%, 41.22\%, 37.46\%, and 36.74\% compared with baseline. This indicates that EGFE can assist front-end developers in improving the readability and maintainability of generated code. This is mainly because the code structure is simplified a lot with less fragmented code snippets. Furthermore, our approach has achieved satisfactory improvements in code availability and maintainability, providing strong evidence that EGFE enhances the industry compliance of GUI-to-code generation. To demonstrate the significance of EGFE, we adopt the Mann-Whitney U test~\cite{fay2010wilcoxon} on four metrics where a $p-value < 0.05$ is considered statistically significant, and a $p-value<0.01$ denotes highly statistically significant. The statistics suggest that EGFE can contribute significantly to the generated code in all four metrics.

\begin{table}[htp]
\centering
\vspace{-0.04in}
\caption{Performance of human evaluation} \label{tab:userstudy}
\resizebox{0.45\textwidth}{!}{
\begin{tabular}{clccl}
\toprule
\multicolumn{1}{c}{\multirow{2}{*}{\textbf{Type}}} & \multicolumn{1}{l}{\multirow{2}{*}{\textbf{Metric}}} & \multicolumn{3}{c}{\textbf{\textbf{Score}}} \\
\cmidrule{3-5}
\multicolumn{1}{c}{}  & \multicolumn{1}{c}{}  & \multicolumn{1}{c}{\textbf{Baseline}} & \multicolumn{1}{c}{\textbf{EGFE}} & \multicolumn{1}{c}{\textbf{P-value}}  \\ 
\midrule
\multirow{4}{*}{Travel}                        
& code availability  &2.52&3.46&$0.0045^{**}$ \\
& modification time &2.60&3.51&$0.0376^{*}$\\
& readability &3.28 &4.17&$0.0493^{*}$\\
& maintainability  &3.10&4.14&$0.0051^{**}$\\ 
\midrule
\multirow{4}{*}{Communication}           
& code availability & 2.36 & 3.49&$0.0022^{**}$\\
& modification time & 2.47 & 3.42&$0.0065^{**}$\\
& readability  &  2.87 & 3.93&$0.0172^{*}$\\
& maintainability &2.42&3.34&$0.0376^{*}$\\ 
\midrule
\multirow{4}{*}{Shopping}                   
& code availability &2.11 &3.29 &$0.0191^{*}$\\
&  modification time &2.29 &3.45&$0.0091^{**}$\\
& readability &2.58 &3.89&$0.0046^{**}$ \\
& maintainability &2.40&3.34&$0.0376^{*}$\\ 
\midrule
\multirow{4}{*}{\textbf{Average}}                      
& code availability &2.33 & 3.41&$0.0321^{*}$\\
& modification time  &2.45 &3.46&$0.0199^{*}$\\
& readability &2.91 &4.00&$0.0029^{**}$\\
& maintainability &2.64 &3.61&$0.0013^{**}$\\
\bottomrule
\end{tabular}
}
\begin{tablenotes}
    \footnotesize
    \item ** denotes $p < 0.01$ and * denotes $p < 0.05$
\end{tablenotes}
\vspace{-0.11in}
\end{table}

\subsection{User Interview}

We hold an interview with these participants. They are generally positive about EGFE and the code improvement it makes. When we questioned the difference they observe between the code changes, P1 indicated that \textit{the code structure after applying EGFE is much clean and easy to understand}. P2 responded by saying, \textit{``This code (applied EGFE) is much close to the version (that) I will submit in my project."} Participants also discussed their experiences with GUI-to-code generation tools i.e., Imgcook, noting that \textit{``I would be wary of using automatic tools, especially for some complex layout"}(P5) or \textit{``(The tools) save my time, but sometimes I have to come back to check design prototypes to manually merge the fragmented elements after noting anomaly in the code"}(P1). We also inquired about their intention to use our EGFE, and most participants replied with positive responses such as \textit{``Definitely (I) will try. It seems able to pull me out from implementing repetitive code again and again"}(P4).

\begin{figure}[thp]
    \centering
    \includegraphics[width=0.43\textwidth]{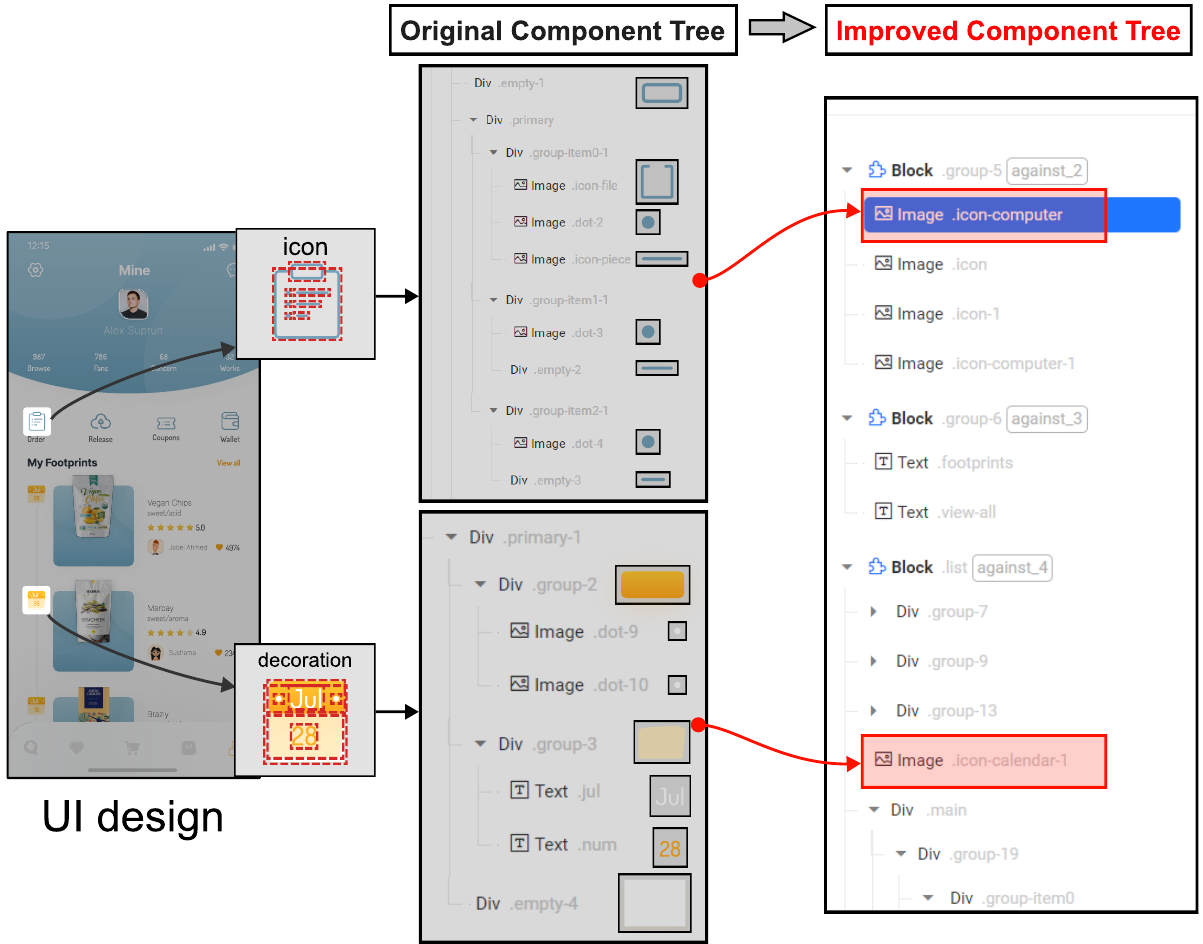}
    \caption{Examples of qualitative analysis with a component tree of generated GUI code}
    \label{fig_case}
    \vspace{-0.24in}
\end{figure}

\subsection{Qualitative Analysis}

To offer a comprehensive evaluation of our EGFE approach, we demonstrate the code generation results with and without our fragmented elements grouping method, using cases from web resources\footnote{\url{https://www.sketchappsources.com/search_dining.html}}. The results are presented in Imgcook online web, as shown in Figure \ref{fig_case}. To better demonstrate the code structure changes, we compare the improvement by our approach based on the component tree rather than specific code. In general, the component tree generated by EGFE tends to be concise and more structured than the original version. For the icon component, the original code mistakenly generates 8 separate image views, resulting in a complicated structure. In contrast, the code generated by EGFE has only one holistic image view representing the icon in GUI, indicating that our approach can understand the GUI layout by encapsulating element grouping information. For the decoration, the original code has redundant UI components, as the code automation software is unaware of the hierarchical information about the input UI design prototype. However, in our approach, the fragmented group is represented by a single image view, making it easier for developers to implement changes and conduct code reviews. As demonstrated in the two cases, structural elements grouping is a crucial step to generate high-quality and maintainable code. Our approach streamlines the code generation process, resulting in less redundant and more reusable GUI code.

\section{Threats to Validity}
There are two main threats to the validity of our approach. The first threat to validity is that our method mainly solves the problem of fragmented elements in the UI design prototypes of mobile application layouts. We did not verify our method on the UI design prototypes that are designed for desktop or web. We will further verify the robustness of the proposed method in future work.

The second threat to validity is the dataset. Although we have performed strict quality control on the dataset annotation process, there may still be some annotation errors. This may reduce the generalization ability of the model. The datasets used by related works were not available due to their company's confidentiality measures, so we were unable to verify the performance of the method on their datasets. To mitigate this issue, we need to obtain UI design prototypes from different sources to verify the generalization ability of the method.
\section{Related Work}
\label{sec:related_work}
\subsection{UI Elements Grouping}

UI grouping is an essential step toward UI understanding. To our knowledge, there are several types of UI grouping: section-level grouping~\cite{Xie2022PsychologicallyInspiredUI}, component-level grouping~\cite{uied, ScreenRecognition}, and element-level grouping~\cite{UILM, uldgnn}. Section-level grouping refers to the division of UI based on human perception, such as grouping components into sections like cards, lists, tabs, and menus. For example, \cite{Xie2022PsychologicallyInspiredUI} proposed a psychologically-inspired perceptual group. They aim to group UI components with similar structures (such as menus, multi-tabs, and cards) based on the Gestalt theory of perception. Component-level grouping is based on identifying UI components with independent semantics to complete downstream tasks~\cite{ScreenRecognition}. For example, recognizing UI components would allow screen readers~\cite{talkback,apple} to provide higher-order perceptual UI structure (e.g. components) for navigating visually impaired users to access applications. REMAUI ~\cite{7372013} infers three Android-specific layouts (LinearLayout, FrameLayout, and ListView) based on hand-craft rules to group widgets.

Fragmented elements grouping (FEG) belongs to element-level grouping. FEG differs from other types in two aspects: granularity and the data to be understood. Firstly, in terms of granularity, the grouping of fragmented elements addresses the smaller granularity of elements within the current UI domain. The fragmented elements are basic shape elements that need to be combined together to express visual semantics. Secondly, in terms of data to be understood, FEG utilizes design prototypes as inputs, whereas other types mainly use images. Therefore, two challenges arise. The first challenge is the detection of UI fragmented elements at a smaller granularity. The second challenge lies in data understanding. FEG not only requires visually locating semantically related fragmented elements but also identifying them within the view hierarchy. UILM~\cite{UILM} is a representative FEG method based on visual learning while requiring a view hierarchy of the design prototype to locate each fragmented element. However, due to the limitations of hand-craft or heuristics rules, this method has poor generalization. To build the relationship between detected UI objects, ~\cite{uldgnn} proposed a graph neural network to learn the structure information in the GUI. However, such a relationship graph cannot represent complex UI element relations in fragmented groups. Furthermore, these methods do not take full advantage of the multimodal features in design prototypes.

\subsection{UI Features Extraction}

The UI image is considered the most prominent feature in UI-related tasks because of its rich sources (e.g., screenshots from real GUIs~\cite{7372013}, sketches and wireframes of design~\cite{Robinson2019Sketch2codeGA}) and the SOTA performance of today's image feature network like Resnet~\cite{7780459} and ViT~\cite{dosovitskiy2020vit}.  With the widespread use of UI images, object detection methods are introduced into UI tasks extensively, especially in UI element detection~\cite{9286056} and UI component recognition~\cite{Zang2021MultimodalIA}. However, images are usually processed flatly while leaving out instructional information in the original UI designs, for example, the description of each component. In this case, the view hierarchy of designs is introduced. Multi-modal solutions~\cite{10.1145/3242587.3242650, Zang2021MultimodalIA, actionbert} with both image features and extra information from the view hierarchy have boosted the performance. In this work, we fully utilize the multi-head feature space mapping and non-local aggregating from multi-head attention and make our approach more powerful by using multi-modal features.

\subsection{GUI-to-Code Generation}
Existing methods for GUI-to-code generation often suffer from code redundancy and code structure complexity. Image-captioning-based GUI-to-code methods~\cite{Beltramelli2018pix2codeGC, Robinson2019Sketch2codeGA, 8453135} are completely unaware of GUI structure during the code generation process. As a result, the generated GUI code is highly redundant for some UI components. For example, the background component in Figure \ref{fig:fragmented_examples} will generate three pieces of the image view, one for each of the fragmented elements. This type of generated code is nothing like the real GUI code developers write. Therefore, it has little practicality. To overcome this limitation, some companies have proposed more mature code generation schemes~\cite{imagecook, codefun}. However, they also hardly recognize fragmented elements very well due to the complex UI view hierarchy as shown in Figure \ref{fig:fragmented_examples}. Without grouping these elements, the generated GUI layouts and widgets have no connection to the corresponding parts in the GUI image. It would be hard to understand how the generated code implements the GUI. With the support of our fragmented elements grouping, GUI-to-code generation can translate the fragmented groups into holistic image views during the code generation process and produce much less redundant and more reusable GUI code.

\section{Conclusion}
\label{sec:conclusion}
In this paper, we propose EGFE, a novel end-to-end approach for recognizing fragmented element groups in UI design prototypes. EGFE effectively leverages multimodal features of UI design prototypes, enabling enhanced UI understanding. To the best of our knowledge, EGFE represents the first automatic end-to-end UI grouping approach for fragmented elements. The evaluation results show that our approach significantly outperforms all other baselines on UI design datasets. An empirical study also confirms the generated code by EGFE tends to be more readable, well-structured, and useful. The positive feedback from participants serves as strong evidence of the validity and potential of EGFE. Our approach addresses the gap in visual intelligence between current GUI-to-code generation techniques, which rely on hand-crafted rules, and the comprehensive GUI-to-code generation at the UI component level. In future work, we plan to incorporate additional UI semantic information and explicitly encode structural details into our model to further enhance its performance.

\begin{acks}
    We would like to thank anonymous reviewers for their insightful and constructive comments, which significantly improve the quality of this paper. This work was supported by the National Key R\&D Program of China (Grant No.2022YFB3303300), the National Natural Science Foundation of China (Grant No.62207023), the Ng Teng Fong Charitable Foundation in the form of ZJU-SUTD IDEA (Grant No.188170-11102), and Alibaba-Zhejiang University Joint Research Institute of Frontier Technologies.
    
\end{acks}

\bibliographystyle{ACM-Reference-Format}
\bibliography{bibsample}

\end{document}